\documentclass[aps,prb,superscriptaddress,amsmath,amssymb,floatfix,twocolumn]{revtex4-1}
\usepackage{times}
\usepackage{graphicx}
\usepackage{subfigure}
\usepackage{color}
\newcommand{\kk}{\mathbf{k}}
\newcommand{\beq}{\begin{equation}}
\newcommand{\eeq}{\end{equation}}
\newcommand{\bdelta}{\boldsymbol{\delta}}

\begin{document}

\title{Spin Dynamics of a $\mathbf{J}_\mathbf{1}$-$\mathbf{J}_\mathbf{2}$-$\mathbf{K}$ Model for the Paramagnetic Phase of Iron Pnictides}

\author{Rong Yu}
\affiliation{Department of Physics and Astronomy, Rice University, Houston,
TX 77005}
\author{Zhentao Wang}
\affiliation{Department of Physics and Astronomy, Rice University, Houston,
TX 77005}
\author{Pallab Goswami}
\affiliation{National High Magnetic Field Laboratory and Department of Physics, Florida State University, Tallahassee, FL 32306}
\author{Andriy Nevidomskyy}
\affiliation{Department of Physics and Astronomy, Rice University, Houston,
TX 77005}
\author{Qimiao Si}
\affiliation{Department of Physics and Astronomy, Rice University, Houston,
TX 77005}
\author{Elihu Abrahams}
\affiliation{Department of Physics and Astronomy, University
 of California Los Angeles, Los Angeles, California 90095}

\begin{abstract}
We study the finite-temperature
spin dynamics of the paramagnetic phase of iron pnictides within an antiferromagnetic
 $J_1-J_2$ Heisenberg model on a square lattice with
 a biquadratic coupling $-K (\mathbf{S}_i \cdot \mathbf{S}_j)^2$ between the nearest-neighbor spins.
Our focus is  on the paramagnetic phase in the parameter regime
of this $J_1-J_2-K$ model where the ground state is
a $(\pi,0)$ collinear antiferromagnet.
We treat the
biquadratic interaction via a Hubbard-Stratonovich decomposition,
and study the resulting effective quadratic-coupling model
using both MSW and SBMF theories;
the results for the spin dynamics derived from the two methods are very similar.
We show that the
spectral weight
of dynamical
structure factor $\mathcal{S}(\mathbf{q},\omega)$
is peaked at ellipses in the momentum space at low excitation energies.
With increasing energy,
the elliptic features expand towards the zone boundary,
and gradually
split into two parts,
forming a pattern around $(\pi,\pi)$. Finally, the spectral weight is anisotropic,
being larger along the major axis of the ellipse than along its minor axis.
These characteristics of the dynamical
structure factor are consistent with the recent measurements of the
inelastic neutron scattering spectra on BaFe$_2$As$_2$
and SrFe$_2$As$_2$.
\end{abstract}

\maketitle

\section{Introduction}
The emergence of superconductivity in iron pnictides~\cite{Kamihara_FeAs,Zhao_Sm1111_CPL08} near
an antiferromagnetically ordered state~\cite{Cruz} in the phase diagram
suggests strong interplay between the superconductivity and magnetism in these materials.
Elucidating the magnetic excitations is therefore important for understanding not only the overall microscopic physics
of these systems but also their superconductivity.
In the parent compounds, the observed$(\pi,0)$ antiferromagnetic order
arises
either within a weak-coupling approach invoking
a Fermi surface nesting,\cite{Graser,Ran,Knolle}
or from a strong-coupling approach whose starting point is a
local moment $J_1-J_2-K$ model.\cite{Si,Yildirim,Ma,Fang:08,Xu:08,Si_NJP,Dai,Uhrig}

The strong-coupling approach is based on the proximity of
the metallic ground state of the parent pnictides to a Mott localization
transition, which gives rise to quasi-local magnetic moments.\cite{Si,Si_NJP,Haule,Kutepov10}
This incipient Mott picture corresponds to a ratio of
$U$ (a measure of the Coulomb repulsions and Hund's couplings among the
Fe 3$d$ electrons) to $t$ (the characteristic bandwidth of the Fe 3$d$ electrons) which
 is not too far below
the Mott threshold $U_c/t$, which is usually of order unity.
This is supported by many experimental observations.
For instance,
the room-temperature electrical resistivity of parent iron pnictides
is so large (even when the residual
resistivity is relatively small signaling the smallness of elastic scattering) that the extracted
mean-free path of quasiparticles would be comparable to the Fermi wavelength;
 this is typical of bad metals near a Mott trasition.
Similarly, the Drude weight in optical conductivity \cite{Qazilbash,Hu} is strongly suppressed from
 its non-interacting counterpart, providing a direct measure of the proximity to the Mott transition.
 This is further corroborated by the
 the temperature-induced spectral weight transfer,\cite{Hu,Yang,Boris} which is also characteristic
 of metals near a Mott transition.
In the spin sector,
zone boundary spin waves
have been observed by inelastic neutron scattering (INS) measurements
in the magnetically ordered state
of several 122 iron pnictides compounds.\cite{Zhao}
Both the large spectral weight and the relatively-small spin damping suggest quasi-localized moments, which
are expected near the Mott transition where the spin excitations arise out of incoherent electronic excitations.
Additional evidence for the incipient Mott transition picture has come from
 the observation of a Mott insulating phase
 in the iron oxychalcogenides.\cite{Zhu} This material contains an expanded Fe square lattice
 compared to the iron pnictides, which reduces $t$, thereby enhancing $U/t$ beyond $U_c/t$ (Ref.~\onlinecite{Zhu}).
 Likewise, the Mott insulating behavior of the alkaline iron selenides \cite{MFang} can also be
 interpreted as the result of a reduced effective $t$ and, correspondingly, an enhanced $U/t$ beyond $U_c/t$.\cite{Yu11,Zhou11}

In the vicinity of $U_c/t$, where correlations are strong, it is natural that the spin Hamiltonian contains not only two-spin
interactions,
such as $J_1$ and $J_2$ Heisenberg exchange between nearest- and next-nearest- neighbor spins on a square
lattice, but also interactions involving higher number of spins. These naturally include, for instance, the ring-exchange coupling involving four spins on a plaquette, and the biquadratic coupling of the form $-K (\mathbf{S}_i \cdot \mathbf{S}_j)^2$ in systems with spin size $S\geqslant 1$.\cite{Fazekas} The subject of the present study is to show how such non-Heisenberg interactions, particularly the biquadratic interaction, influence the spin dynamics in the paramagnetic phase.

Spin dynamics in the parent iron pnictides have been most extensively studied in the
low-temperature state ($T<T_N$)  with both antiferromagnetic order and orthorhombic structural distortion.
Here, the INS experiments up to high energies (on the order of $200$ meV)
show that the spin wave excitations in these compounds are highly anisotropic,
with a dispersion which can be understood in terms of
an anisotropic $J_{1x}-J_{1y}-J_2$ model with $J_{1x}\neq J_{1y}$.\cite{Zhao,ZhaoSr,Han09}.
The anisotropy in the nearest-neighbor coupling is compatible with the orthorhombic
structure, and its degree could reflect an orbital ordering.\cite{Singh,Jaanen,Lv, AN}
Detailed theoretical studies of the magnetic excitations in the ordered phase have been
carried out in such a $J_{1x}-J_{1y}-J_2$ model,\cite{Applegate} and in
a $J_1-J_2-K$ model.\cite{WysockiNP11,Stanek11} It should also be noted that terms such as the biquadratic coupling could be inferred from the sublattice
angle dependence of the ground-state energy in LSDA calculations~\cite{Yaresko},
and were shown to appear naturally as a result of the orbital ordering
between Fe d$_{xz}$ and d$_{yz}$ orbitals~\cite{AN}.

Our focus is instead on the spin dynamics in the {\em paramagnetic phase} of the parent iron pnictides,
which has only recently been studied experimentally.
The initial work by Diallo {\it et al.}\cite{Diallo} measured the spin dynamics of CaFe$_2$As$_2$
at relatively low energies, below 70 meV.
Theoretically, four of us\cite{Goswami11} studied the spin dynamics in the paramagnetic phase of the
$J_1-J_2$ model (with or without an additional fermion damping).
We showed that the experimentally observed elliptical features of the spin spectral weight
in momentum space are well-described by this model and we determined the change
to the elliptical features at high energies.

More recently, Harriger {\it et al.}\cite{Harriger}
reported measurements of the spin dynamics in the paramagnetic phase up to high energies
(above 200 meV) in BaFe$_2$As$_2$. The INS measurements confirmed the quasi-two-dimensional
spin dynamics found at low energies,\cite{Diallo} and characterized the
evolution of the low-energy elliptic features as they expand towards the zone boundary as
the energy is raised, and determined the high-energy dispersion which appears to require
a $J_{1x}\neq J_{1y}$ description even though the paramagnetic phase has a \emph{tetragonal} structure.
 Similar data have also been
reported by Ewings {\it et al}.\ in SrFe$_2$As$_2$.\cite{Ewings}
Theoretically, Park {\it et al.}\cite{ParkHauleKotliar11}  analyzed the spin dynamics in the paramagnetic state
within a dynamical mean-field theory (DMFT) for interactions
$U/t \lessapprox U_c/t$, demonstrating that
the
DMFT approach captures key features of the neutron scattering results,
including the ellipticity of the map of the structure-factor peak in the Brillouin zone.

In this paper, we study the spin dynamics of the $J_1-J_2-K$ model in the tetragonal paramagnetic
phase using both modified spin wave (MSW) and Schwinger boson mean-field (SBMF) theories.
The results from the two methods are in very good quantitative agreement with each other.
We show that, for a moderate biquadratic coupling $K$,
the dynamical structure factor $\mathcal{S}(\mathbf{q},\omega)$ has not only
elliptic features near $(\pi,0)$,
which expand with increasing energy and split into peaks surrounding $(\pi,\pi)$,
but also an anisotropic distribution of the spectral weight that is larger along the
major axis of each ellipse than along its minor axis. These properties agree well
with the INS experiments  \cite{Harriger,Ewings}

The remainder of the paper is organized as follows. In Sec.~\ref{Sec:Model} we introduce
the $J_1-J_2-K$ model and describe the MSW and SBMF
theories used in this paper. In Sec.~\ref{Sec:PhD} we show how the biquadratic coupling
$K$ influences the mean-field phase diagram and magnetic excitation spectrum.
In Sec.~\ref{Sec:Sqw} we calculate the dynamical structure factor $\mathcal{S}(\mathbf{q},\omega)$.
We also show that the spectral weight exhibits anisotropic features,
discuss the evolution of the anisotropic features with increasing excitation energy, and explain
how these properties arise
from our theory. In Sec.~\ref{Sec:Disc} we first discuss some possible generalizations of the
 $J_1-J_2-K$ model we are studying in this paper. In the same section, we then consider the effect
 of itinerant electrons, and compare our
study with other theoretical approaches to the spin dynamics. Sec.~\ref{Sec:expt} is devoted to a comparison
with the INS experiments on the paramagnetic phases of the parent 122 iron pnictides,
and Sec~\ref{Sec:Conclusion} contains a few concluding remarks. In three appendices,
we expound on the Ising transition at small $J_1/J_2$ ratios,
and discuss the effects
of both the ring-exchange interactions
and interlayer exchange couplings.

\section{Model and Methods}\label{Sec:Model}
The  $J_1-J_2-K$ model is defined on a two-dimensional (2D) square lattice with the following Hamiltonian:
\begin{eqnarray}\label{Eq:J1J2KHam}
H &=& J_1\sum_{i,\boldsymbol{\delta}}\mathbf{S}_{i}\cdot \mathbf{S}_{i+\boldsymbol{\delta}}
+J_2\sum_{i,\boldsymbol{\delta}^\prime}\mathbf{S}_{i}
\cdot \mathbf{S}_{i+\boldsymbol{\delta}^\prime}
\nonumber\\
&& - K \sum_{i,\boldsymbol{\delta}} \left(\mathbf{S}_{i}\cdot \mathbf{S}_{i+\boldsymbol{\delta}}\right)^2,
\end{eqnarray}
where $J_1$ and $J_2$ respectively denote
the antiferromagnetic exchange couplings between spins
located in the nearest neighbor ($\boldsymbol{\delta}=\hat{x}$,$\hat{y}$) and
next-nearest neighbor ($\boldsymbol{\delta}^\prime=\hat{x}\pm\hat{y}$)
sites. $K$ is the coupling for the biquadratic interaction between the nearest neighbor spin pairs.

To fully explain the experimentally observed $(\pi,0,\pi)$ antiferromagnetic order, an exchange coupling
along the third dimension, $J_z$ should also be included. However, we find the model defined
in Eq.~\eqref{Eq:J1J2KHam} already allows us to understand the experimentally observed quasi-2D spin dynamics.
Hence, we concentrate on this 2D model in the main text, and discuss the influence of the interlayer coupling
$J_z$ on the spin dynamics in Appendix~\ref{Sec:PhD3D}.

The Hamiltonian of Eq.~\eqref{Eq:J1J2KHam} is studied using both MSW\cite{Takahashi1,Takahashi2}
and SBMF\cite{Arovas} methods.
Here, we focus on the parameter regime where the ground state has a collinear $(\pi,0)$ antiferromagnetic order,
and decompose the biquadratic interaction term of the Hamiltonian using two Hubbard-Stratonovich
fields $\Gamma_{i,\hat{x}(\hat{y})}$. The effective Hamiltonian
reads as
\begin{eqnarray}\label{Eq:HSHam}
H &=& J_1\sum_{i,\boldsymbol{\delta}}\mathbf{S}_{i}\cdot \mathbf{S}_{i+\boldsymbol{\delta}}
+J_2\sum_{i,\boldsymbol{\delta}^\prime}\mathbf{S}_{i}
\cdot \mathbf{S}_{i+\boldsymbol{\delta}^\prime} 
\nonumber\\
 & & - 2K \sum_{i,\boldsymbol{\delta}} \Gamma_{i,\boldsymbol{\delta}}
 \mathbf{S}_{i}\cdot \mathbf{S}_{i+\boldsymbol{\delta}} + K\sum_{i,\boldsymbol{\delta}}
 \Gamma_{i,\boldsymbol{\delta}}^2.
\end{eqnarray}
At the mean-field level, the Hubbard-Stratonovich fields are treated as static quantities,
and can be expressed using equal-time spin correlators as:
$\Gamma_{i,\boldsymbol{\delta}}=\langle \mathbf{S}_i\cdot\mathbf{S}_{i+\boldsymbol{\delta}}\rangle$.
The Hubbard-Stratonovich transformation itself is exact.
The static approximation is made in accordance with the level of approximation inherent
to the MSW and SBMF methods, which incorporate static self-energies for the respective boson fields.
As shown below, our approach has two important features:
i) it is capable of studying the Ising correlations at nonzero temperatures;
and ii) the MSW and SBMF approaches yield consistent results.

\subsection{The modified spin wave theory}\label{Sec:Model-MSW}
The MSW theory~\cite{Takahashi1,Takahashi2} has been applied to the $J_1-J_2$ model
by four of us.\cite{Goswami11} In this approach, a local spin quantization axis is defined at each site
along the classical ordering direction $\mathbf{\Omega}_{i}^{cl}$. The Hamiltonian in
Eq.~\eqref{Eq:HSHam} is then expressed in terms of Dyson-Maleev (DM) bosons via a local DM
transformation: $\mathbf{S}_i\cdot\mathbf{\Omega}_i^{cl}=S-a^\dagger_i a_i$,
$\mathbf{S}^+_i=\sqrt{2S}(1-a^\dagger_i a_i/2S) a_i$, and $\mathbf{S}^-_i=\sqrt{2S}a^\dagger_i$.
Minimizing the free energy under the constraint of zero sublattice magnetization $\langle S-a^\dagger_ia_i\rangle=0$
by introducing a
Lagrange multiplier
$\mu$, and with respect to $\Gamma_{\delta}$ ($=\Gamma_{i,\boldsymbol{\delta}}$,
by assuming translational symmetry), we obtain
\begin{eqnarray}\label{Eq:Gammas}
 \Gamma_{x} &=& \cos^2\frac{\phi}{2}f^2_x -\sin^2\frac{\phi}{2}g^2_x, \nonumber\\
 \Gamma_{y} &=& \sin^2\frac{\phi}{2}f^2_y -\cos^2\frac{\phi}{2}g^2_y,
\end{eqnarray}
where $\phi=\arccos (\mathbf{\Omega}_i^{cl}\cdot\mathbf{\Omega}_{i+\hat{x}}^{cl})$.
$f_{\delta}=\langle a^\dagger_i a_{i+\boldsymbol{\delta}}\rangle$ and
$g_{\delta}=\langle a_i a_{i+\boldsymbol{\delta}}\rangle$ are the ferromagnetic and
antiferromagnetic bond operators, respectively. Minimizing the free energy with respect to $\phi$ gives
either $\sin\phi=0$ for nonzero $f_\delta$ and $g_\delta$, or $\phi$ can be arbitrary if $f_\delta=g_\delta=0$.
This defines two phases separated by a mean-field temperature scale~\cite{Goswami11} $T_{\sigma0}$:
at $T>T_{\sigma0}$, $\phi$ is arbitrary, and the system has $C_{4v}$ lattice rotational symmetry;
while for $T<T_{\sigma0}$, the $C_{4v}$ symmetry is broken and the system is Ising ordered,
corresponding to either $\phi=0$ or $\phi=\pi$. In MSW theory, the Ising order parameter
can be defined as $\sigma=2(\cos^2\frac{\phi}{2}(f^2_x+g^2_y)-\sin^2\frac{\phi}{2}(f^2_y+g^2_x))$.
From Eq.~\eqref{Eq:Gammas}, if we define $\Gamma_{\pm} =( \Gamma_x\pm\Gamma_y)/2$
as the symmetric and antisymmetric Hubbard-Stratonovich fields, we find that $\Gamma_-=\sigma/4$.

Minimizing the free energy with respect to other variational parameters, we obtain a set of self-consistent equations:
\begin{eqnarray}
 &&f_{\delta}=m_0+\frac{1}{\mathcal{N}}{\sum_{\mathbf{k}}}^\prime\frac{B_{\mathbf{k}}}
{\varepsilon_{\mathbf{k}}}\left(n_{\mathbf{k}}+\frac{1}{2}\right)\cos
(\mathbf{k}\cdot\boldsymbol{\delta}), \label{Eq:MSWSelf-Consist1}\\
&&g_{\delta}=m_0+\frac{1}{\mathcal{N}}{\sum_{\mathbf{k}}}^\prime\frac{A_{\mathbf{k}}}
{\varepsilon_{\mathbf{k}}}
\left(n_{\mathbf{k}}+\frac{1}{2}\right)\cos (\mathbf{k}\cdot\boldsymbol{\delta}^\prime), \label{Eq:MSWSelf-Consist2}\\
&&S+\frac{1}{2}=m_0+\frac{1}{\mathcal{N}}{\sum_{\mathbf{k}}}^\prime\frac{B_{\mathbf{k}}}
{\varepsilon_{\mathbf{k}}}\left(n_{\mathbf{k}}+\frac{1}{2}\right), \label{Eq:MSWSelf-Consist3}
\end{eqnarray}
where $\mathcal{N}$ is the total number of lattice sites,
${\boldsymbol{\delta}}=\hat{x},\hat{y}$, and ${\boldsymbol{\delta}^\prime} = \hat{x},\hat{y},\hat{x}\pm \hat{y}$.
In Eqs.~\eqref{Eq:MSWSelf-Consist1}-\eqref{Eq:MSWSelf-Consist3},
\begin{eqnarray}
 A_{\mathbf{k}} &=& 2\sin^2\frac{\phi}{2}\tilde{J}_{1x}g_x\cos k_x + 2\cos^2
 \frac{\phi}{2}\tilde{J}_{1y}g_y\cos k_y \nonumber\\
 && + 4J_2g_{x+y}\cos k_x\cos k_y, \label{Eq:DispersionA}\\
 B_{\mathbf{k}} &=& 4J_2g_{x+y}-\mu + 2\sin^2\frac{\phi}{2}(\tilde{J}_{1x}g_x
 - \tilde{J}_{1y}f_y(1-\cos k_y)) \nonumber\\
 && + 2\cos^2\frac{\phi}{2}(\tilde{J}_{1y}g_y
 -\tilde{J}_{1x}f_x(1-\cos k_x)) , \label{Eq:DispersionB}
\end{eqnarray}
and the Bogoliubov angle $\theta_{\mathbf{k}}$ is defined via $\tanh 2\theta_\mathbf{k}= A_\mathbf{k}/B_\mathbf{k}$.
The boson dispersion $\varepsilon_{\mathbf{k}}=\sqrt{B^2_\mathbf{k}-A^2_\mathbf{k}}$ and the boson number
$n_{\mathbf{k}}=[\exp(\varepsilon_{\mathbf{k}}/k_BT)-1]^{-1}$. At $T=0$, the spectrum of the DM bosons becomes
gapless at wave vector $\mathbf{Q}$ and $\mathbf{0}$. This corresponds to a long-range
antiferromagnetic order at $\mathbf{Q}\neq0$ with a nonzero spontaneous magnetization $m_0$.
In this case, the summation ${\sum_{\mathbf{k}}}^\prime$ runs over
all $\mathbf{k}$ values that make $\varepsilon_{\mathbf{k}}>0$, and the contribution from
the $\varepsilon_{\mathbf{k}}=0$ terms
is taken into account separately by $m_0$. For $T>0$, $m_0=0$, and the system is paramagnetic.
Here the summation
is performed in the full momentum space.
In the presence of a small third-dimension coupling $J_z$, there will be a nonzero mean-field
N\'{e}el
temperature, $T_{N0}$; this is discussed in Appendix~\ref{Sec:PhD3D}.

Note that these self-consistent equations are exactly the same as those for the
isotropic $J_1-J_2$ model.\cite{Goswami11} But the definitions of $A_{\mathbf{k}}$ and $B_{\mathbf{k}}$ are different.
In Eqs.~\eqref{Eq:DispersionA} and \eqref{Eq:DispersionB} above, we defined the effective exchange couplings $J_{1x}$ ($J_{1y}$) along the $x(y)$ direction as follows:
\begin{equation} \label{Eq:Jxy}
\tilde{J}_{1x(y)}=J_1-2K\Gamma_{x(y)},
\end{equation}
expressed in terms of the Hubbard-Stratonovich fields $\Gamma_{x(y)}$ of spin-spin correlators in Eq.~(\ref{Eq:Gammas}).
 Although in the $J_1-J_2-K$ model the bare nearest neighbor exchange coupling $J_1$ is still isotropic, a nonzero
 biquadratic coupling $K$ leads to an anisotropic \emph{effective coupling} $\tilde{J}_{1x}\neq\tilde{J}_{1y}$ in the Ising ordered
 phase where $\Gamma_x\neq\Gamma_y$, {\it i.e.} the nearest-neighbor spin correlators along $x$ and $y$ are unequal,
 similarly to the situation found originally\cite{Chandra} for the $J_1-J_2$ model.

\subsection{The Schwinger boson theory}\label{Sec:Model-SB}

In the Schwinger boson representation,\cite{Arovas} the SU($2$) spin operators are  rewritten in terms
of two Schwinger bosons via the transformation: $S_{i}^{z}=\frac{1}{2}(a_{i}^{\dagger}a_{i}-b_{i}^{\dagger}b_{i})$,
$S_{i}^{+}=a_{i}^{\dagger}b_{i}$, and $S_{i}^{-}=b_{i}^{\dagger}a_{i}$.
To limit the boson Hilbert space to the physical sector, a constraint
$a_i^\dagger a_i + b_i^\dagger b_i = 2S$
is imposed on each site. This can be generalized to the case of either SU($N$)
\cite{ArovasAuerbach} or SP($N$) \cite{ReadSachdev,Flint} spins, in either case there
will be $N$ boson degrees of freedom at each site.
For the experimentally observed $\mathbf{Q}=(\pi,0)$ or $(0,\pi)$ antiferromagnetic collinear
phase in the 122 parent compounds, the $(ab)$-plane spin-spin correlations are expressed as:
\begin{eqnarray}
 \mathbf{S}_{i}\cdot\mathbf{S}_{j} &=&
  -(1-\Theta(i,j)) [2\hat{g}_{ij}^{\dagger}\hat{g}_{ij}-S^{2}]\nonumber\\
  & & +\Theta(i,j) [2\hat{f}_{ij}^{\dagger}\hat{f}_{ij}-S(S+1)],
\end{eqnarray}
where $\hat{f}_{\delta}\!\equiv\! f_{i,i+\boldsymbol{\delta}}\!=\!\frac{1}{2} (a_{i}^{\dagger}a_{i+\boldsymbol{\delta}}
+b_{i}^{\dagger}b_{i+\boldsymbol{\delta}})$
and $\hat{g}_{\delta}\!\equiv\! g_{i,i+\boldsymbol{\delta}}\!=\!\frac{1}{2}
(a_{i}b_{i+\boldsymbol{\delta}}-b_{i}a_{i+\boldsymbol{\delta}})$ are respectively the ferromagnetic
and antiferromagnetic bond operators. The function $\Theta(i,j)=1$ if $i$ and $j$ are on the
same stripe sublattice, and $\Theta(i,j)=0$ if $i$ and $j$ are on different stripe sublattices.
The Hubbard-Stratonovich field is then $\Gamma_{\delta}=|\hat{f}_{\delta}|^{2}-|\hat{g}_{\delta}|^{2}$,
and in the case of $(\pi,0)$ ordering we find
\begin{eqnarray}\label{Eq:Gamma-SB}
 \Gamma_{x} &=& -g^2_x, \nonumber\\
 \Gamma_{y} &=& f^2_y.
\end{eqnarray}
Comparing this to Eq.~(\ref{Eq:Gammas}), we see that the spin correlators coincide with
those in the MSW theory if one sets $\phi=\pi$. Similarly, the case of $(0,\pi)$ ordering
corresponds to $\phi=0$ in Eq.~(\ref{Eq:Gammas}). In both cases, $\Gamma_{x}$ and $\Gamma_{y}$ have opposite sign, leading to the anisotropy in the effective spin-spin exchange couplings $\tilde{J}_{1x} \neq \tilde{J}_{1y}$, from Eq.~(\ref{Eq:Jxy}).

By introducing Fourier transformation~\cite{Ceccatto}
\begin{eqnarray}
a_{i} & = & \frac{1}{\sqrt{\mathcal{N}}} \sum_{\mathbf{k}} a_{\mathbf{k}} e^{i(\mathbf{k}-\frac{\mathbf{Q}}{2})\cdot\mathbf{r}_{i}},\\
b_{i} & = & \frac{1}{\sqrt{\mathcal{N}}} \sum_{\mathbf{k}} b_{\mathbf{k}} e^{i(\mathbf{k}+\frac{\mathbf{Q}}{2})\cdot\mathbf{r}_{i}},
\end{eqnarray}
and making a Bogoliubov transformation to a new quasiparticle creation/annihilation operators
$\alpha_k = \cosh\theta_\kk a_\kk + i \sinh\theta_k b^\dagger_{-\kk}$,
one arrives at the mean-field free energy density, which can be generalized to the Sp($N$) form~\cite{Flint}
\begin{eqnarray}
F_{MF} &=& \frac{NT}{\mathcal{N}}\sum_\kk\ln\left[ 2\sinh\left(\frac{\omega_\kk}{2T}\right)  \right]
+ N\lambda\left(S+\frac{1}{2}\right) \nonumber \\
&-&  \frac{N z}{8} \sum_{\bdelta} J_{\bdelta} (|f_{\bdelta}|^2 - |g_{\bdelta}|^2),
\label{free-SB}
\end{eqnarray}
where $z$ is the coordination number, and $\lambda$ is the Lagrange multiplier
associated with the imposed constraint that on average, the number of bosons per
site $\sum_{\sigma=1}^N n_{i\sigma} = NS$. Here $\omega_\kk = \sqrt{(B_\kk - \lambda)^2 - A_\kk^2}$
is the dispersion of the Bogoliubov quasiparticles, expressed~\cite{Ceccatto} in terms of the variables
\begin{equation}
A_\kk = i\sum_{\bdelta} J_{\bdelta}\,g_{\bdelta}\; e^{-i\kk\cdot\bdelta};\quad  B_\kk
= \sum_{\bdelta} J_{\bdelta}\,f_{\bdelta}\; e^{-i\kk\cdot\bdelta} ;
\end{equation}
the Bogoliubov angle  $\tanh2\theta_{\mathbf{k}}\! =\! A_{\mathbf{k}}/(\lambda-B_{\kk})$.
The dispersion relation $\omega_\kk$ explicitly depends on the ordering wave-vector ${\mathbf Q}$
and has minima around $\kk=\pm\mathbf{Q}/2$.
In the regime when $J_2>J_1/2$, the minimization of the free energy results in  $\mathbf{Q}=(\pi,0)$ or $(0,\pi)$.
For example, for $\mathbf{Q}=(\pi,0)$, the expressions for $A_\kk$
 and $B_\kk$ become:
\begin{eqnarray}
\label{Eq:SBDispEqA}
A_{\mathbf{k}} & = & 2 \tilde{J}_{1x}g_{x}\sin k_{x}
+4J_{2}g_{x+y}\sin k_{x}\cos k_{y} 
,\\
\label{Eq:SBDispEqB} B_{\mathbf{k}} & = & 2 \tilde{J}_{1y}f_{y}\cos k_{y}.
\end{eqnarray}
In the large-$N$ limit of the Sp($N$) spin, the mean-field free energy Eq.~(\ref{free-SB})
becomes exact.\cite{Arovas,Flint} The observable magnetic excitation spectrum is obtained
from $\omega_\kk$ by a $\mathbf{Q}/2$ shift: $\varepsilon_\kk = \omega_{\kk-\mathbf{Q}/2}$.
At $T=0$, the magnetic order results in the gapless Goldstone modes at $\kk=\mathbf{0}$ and $\mathbf{Q}$,
as expected. The SBMF theory is known to reproduce well the
spectrum of spin waves in both ferro- and antiferro-magnets.\cite{Arovas,Ceccatto}

Below, we focus on the paramagnetic phase at $T>0$, with short-range
$\mathbf{Q}=(\pi,0)$ antiferromagnetic correlations (the case
$\mathbf{Q}=(0,\pi)$ is obtained by $C_4$ lattice rotation).
We obtain the following self-consistent equations from the saddle-point minimization of the free energy Eq. (\ref{free-SB}):
\begin{eqnarray}
\label{Eq:SBSelfConsistEq1}
&& f_{\delta} = \frac{1}{\mathcal{N}} \sum_{\mathbf{k}}
\frac{B_{\mathbf{k}}-\lambda}{\omega_{\mathbf{k}}} \left(n_{\mathbf{k}}+\frac{1}{2}\right)
\cos\left(\mathbf{k}\cdot\boldsymbol{\delta}\right), \\
\label{Eq:SBSelfConsistEq2} && g_{\delta^\prime} = \frac{1}{\mathcal{N}}\sum_{\mathbf{k}}
\frac{A_{\mathbf{k}}}{\omega_{\mathbf{k}}} \left(n_{\mathbf{k}}+\frac{1}{2}\right)
\sin\left(\mathbf{k} \cdot\boldsymbol{\delta^\prime}\right), \\
\label{Eq:SBSelfConsistEq3}
&& S+\frac{1}{2} = \frac{1}{\mathcal{N}}\sum_{\mathbf{k}}
\frac{B_{\mathbf{k}}-\lambda}{\omega_{\mathbf{k}}}(n_{\mathbf{k}}+\frac{1}{2}),
\end{eqnarray}
where  $\boldsymbol{\delta}=\hat{y}$, and $\boldsymbol{\delta^\prime}=\hat{x},\hat{x}\pm\hat{y}$.
Under the transformation $B_{\mathbf{k}}-\lambda\rightarrow B_{\mathbf{k}}$ and
$\mathbf{k}\rightarrow \mathbf{k}-\mathbf{Q}/2$, Eqs.~\eqref{Eq:SBDispEqA}-\eqref{Eq:SBSelfConsistEq3}
in the SBMF theory and Eqs.~\eqref{Eq:MSWSelf-Consist1}-\eqref{Eq:DispersionB}
in the MSW mean-field theory have exactly the same form in the short-range $(\pi,0)$
correlated paramagnetic phase.
Therefore, the two methods yield exactly the same mean-field phase diagram and boson dispersion,
as corroborated by explicit numerical comparison. We further verified that these two theories give similar
results for the spin dynamics of the $J_1-J_2-K$ model.

\section{Mean-field phase diagram and excitation spectrum}\label{Sec:PhD}

\begin{figure}[t!]
\centering\includegraphics[scale=0.3]{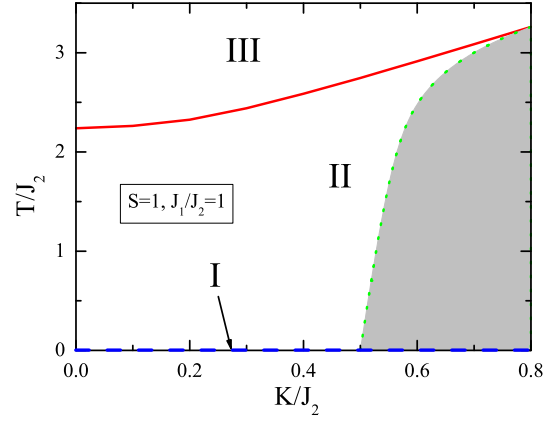}
\caption{(Color online)
Mean-field phase diagram in the MSW theory for $S=1$ and $J_1/J_2=1$.
Phases I, II, and III respectively denote the $(\pi,0)/(0,\pi)$ long-range antiferromagnetically
ordered state (at $T=0$), the Ising ordered paramagnetic state, and the isotropic paramagnetic state.
The solid red curve refers to the mean-field temperature scale $T_{\sigma0}=T_0$. In the shaded region
the effective exchange coupling $\tilde{J}_{1y}<0$.
}
\label{fig:phd2D}
\end{figure}

\begin{figure}[h]
\centering\includegraphics[scale=0.35]{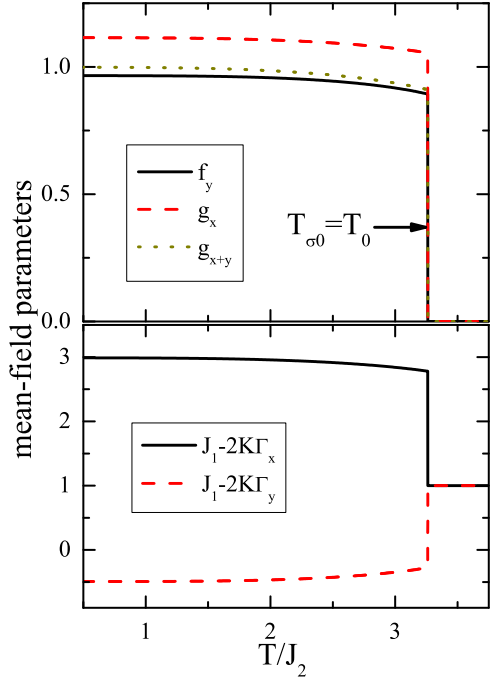}
\caption{(Color online)
The temperature evolution of the mean-field parameters in the MSW theory
for $S=1$, $J_1/J_2=1$, and $K/J_2=0.8$.
}
\label{fig:MFP2D}
\end{figure}

Since INS measurements suggest $J_1\sim J_2$ for several 122 compounds,\cite{Zhao,Diallo,Harriger}
our discussion on the $J_1-J_2-K$ model is focused on this parameter regime.
Fig.~\ref{fig:phd2D} shows the mean-field phase diagram of the 2D $J_1-J_2-K$ model
using the MSW method for $S=1$ and $J_1/J_2=1$. We identify three different phases.
Phase I corresponds to the $(\pi,0)/(0,\pi)$ antiferromagneticaly long-range ordered phase;
it exists only at $T=0$ in the 2D model. Phase II and phase III are both paramagnetic.
They are separated by a mean-field Ising transition temperature $T_{\sigma0}$.
We find that for $J_1/J_2=1$, this transition is first-order, as shown in Fig.~\ref{fig:MFP2D}.
But it can be either first-order or second-order for $J_1/J_2\lesssim 0.9$, as discussed in more detail
in Appendix~\ref{Sec:IsingTransition}.
In the low-temperature phase II, either $f_x\neq f_y$ or $g_x\neq g_y$ (see Fig.~\ref{fig:MFP2D}),
corresponding to an Ising ordered phase with either $(\pi,0)$ or $(0,\pi)$ short-range antiferromagnetic correlations.
This Ising ordered phase already exists in the isotropic $J_1-J_2$ model.\cite{Chandra,Flint,Goswami11}
But here, we find that a nonzero $K$ enhances $T_{\sigma0}$, and $K$ drives the effective
nearest-neighbor exchange couplings to be anisotropic. As shown in Fig.~\ref{fig:MFP2D}, in the $(\pi,0)$
Ising ordered phase (corresponds to $\phi=\pi$), the effective coupling $\tilde{J}_{1y}$ can even be ferromagnetic.
This is important for understanding the experimentally observed anisotropic magnetic excitations
at high energies
in Ca-122\cite{Zhao} and Ba-122.\cite{Harriger} Phase III at $T>T_{\sigma0}$ is the Ising disordered paramagnetic phase.
In this phase the effective nearest-neighbor exchange couplings are isotropic because the nearest-neighbor bond
correlators are zero. But the next-nearest-neighbor bond correlations may still be finite in this phase. One may define
another temperature scale $T_0$, above which the next-nearest-neighbor bond correlations vanish and the system are decoupled into isolated local moments. Note that $T_0$ does not refer to a phase transition, and the discontinuity
of the bond correlations at $T_0$ is an artifact of the mean-field theory.\cite{Takahashi1} In general,
$T_0$ and $T_{\sigma0}$ are two different temperature scales satisfying
$T_0\geqslant T_{\sigma0}$.\cite{Goswami11,Flint} But for $J_1/J_2\gtrsim 0.9$, $T_0=T_{\sigma0}$
for any $K/J_2$ ratio, as shown in Fig.~\ref{fig:MFP2D}. The phase diagram obtained in the SBMF theory is identical to the one shown in Fig.~\ref{fig:phd2D}.

\begin{figure} [h]
\centering\includegraphics[scale=0.3]{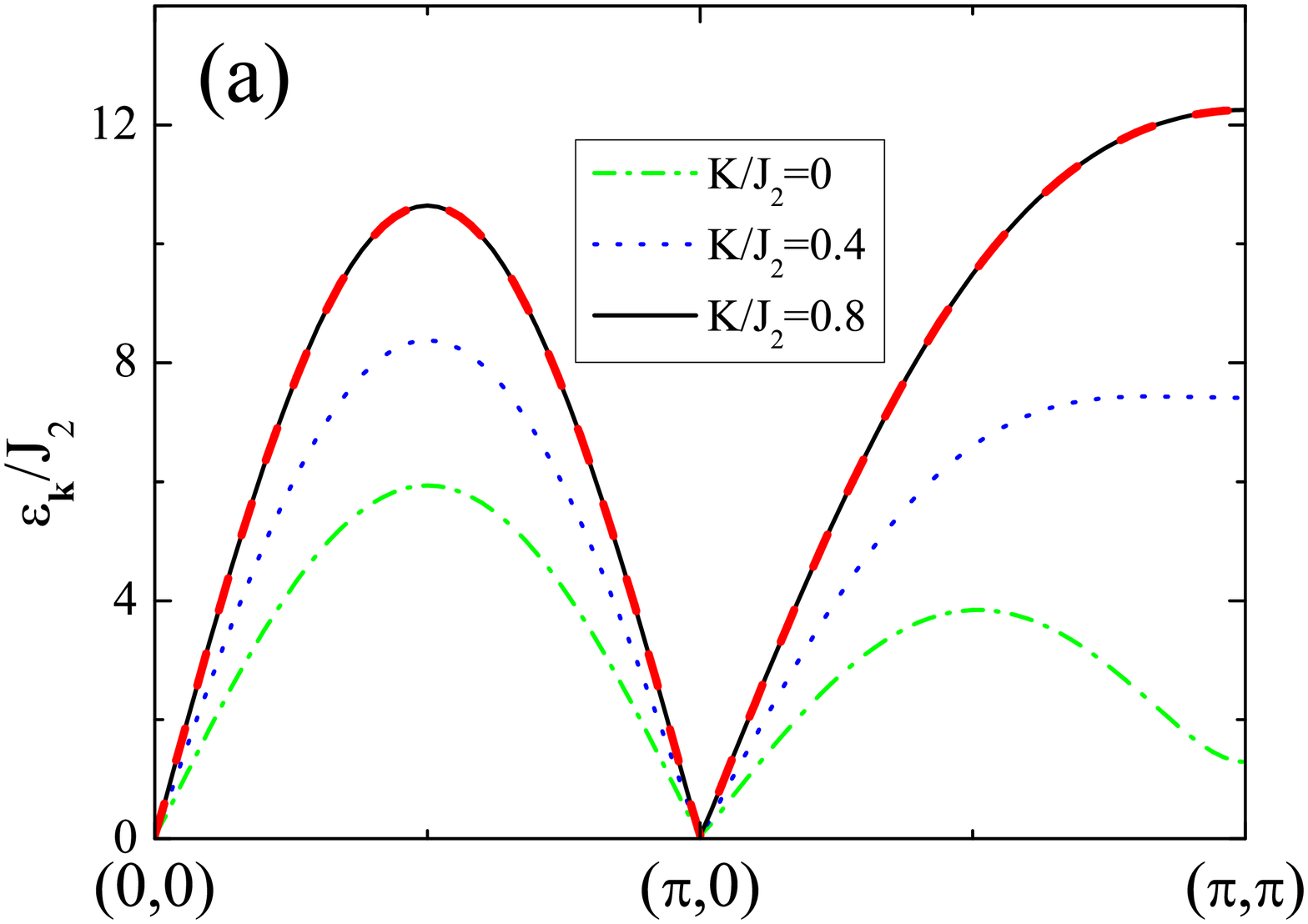}\\ 
\centering\includegraphics[scale=0.3]{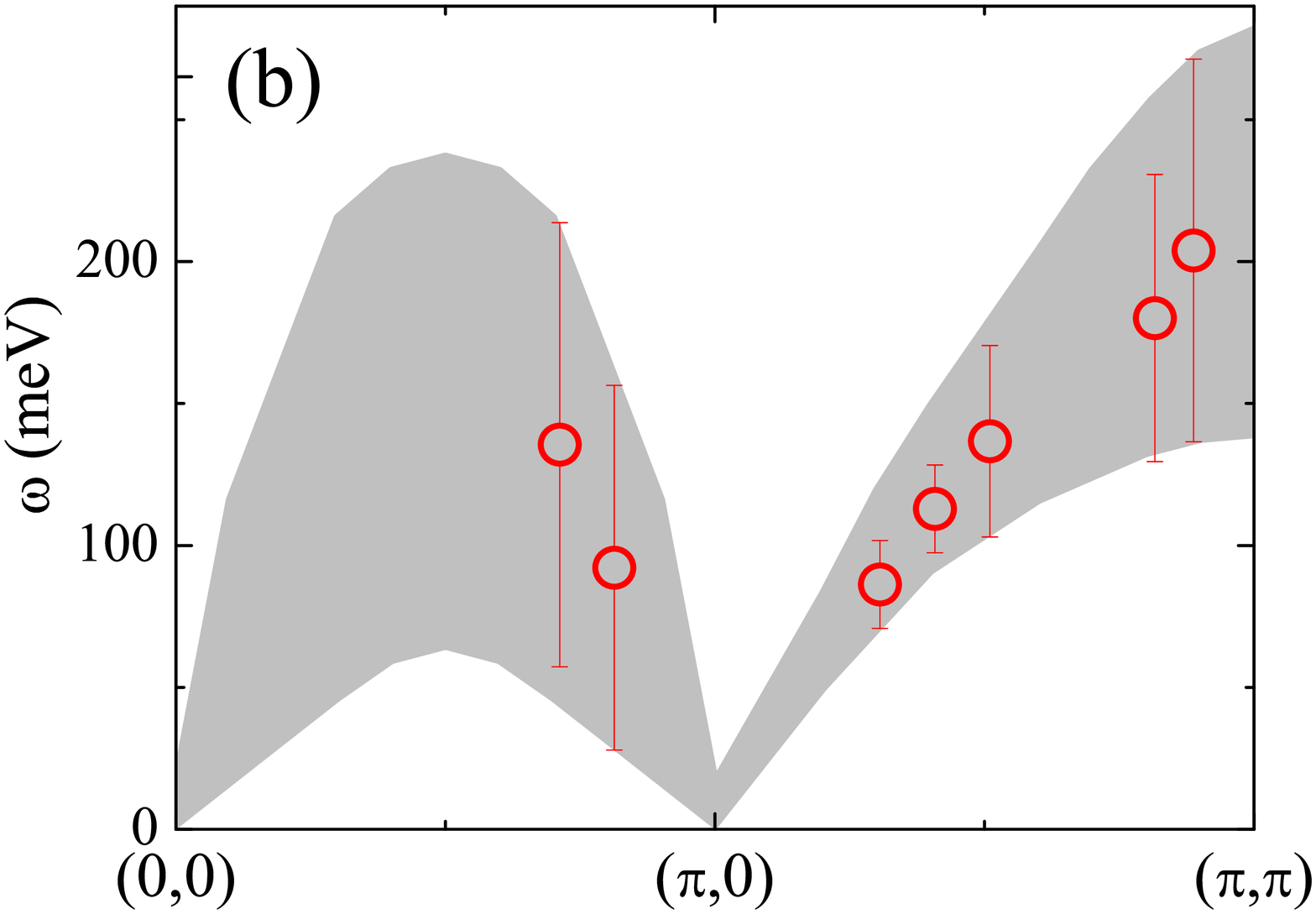} 
\caption{(Color online)
(a): MSW dispersion along high symmetry directions in the
paramagnetic Brillouin zone for the 2D $J_1-J_2-K$ model at $S=1$,
$J_1/J_2=1.0$, $T/J_2=1.0$, and various $K$ values.
For comparison, the red dashed curve shows the dispersion in the
SBMF theory for the same parameters and $K/J_2=0.8$.
The gaps at $(0,0)$ and $(\pi,0)$ are too small to be seen in the figure.
(b): The symbols show the dispersion from the INS data at $T=150$ K in BaFe$_2$As$_2$,
taken from Ref.~\onlinecite{Harriger}.
The data can be fit by any of the theoretical dispersion curves [as in (a)] that lie within the shaded region.
}
\label{fig:Dispersions}
\end{figure}

The finite coupling $K$ not only changes the phase boundary of the mean-field phase diagram,
but can also dramatically influence the boson excitation spectrum. In Fig.~\ref{fig:Dispersions}(a),
we show the dispersions of the DM and Schwinger bosons along two high-symmetry directions
in momentum space for various $K$ values in phase II with $\phi=\pi$ using the same
parameters as in Fig.~\ref{fig:phd2D}. We see that the dispersion in Schwinger boson theory
matches the one in the MSW theory exactly.

\begin{figure} [h]
\centering
\includegraphics[
width=85mm
]{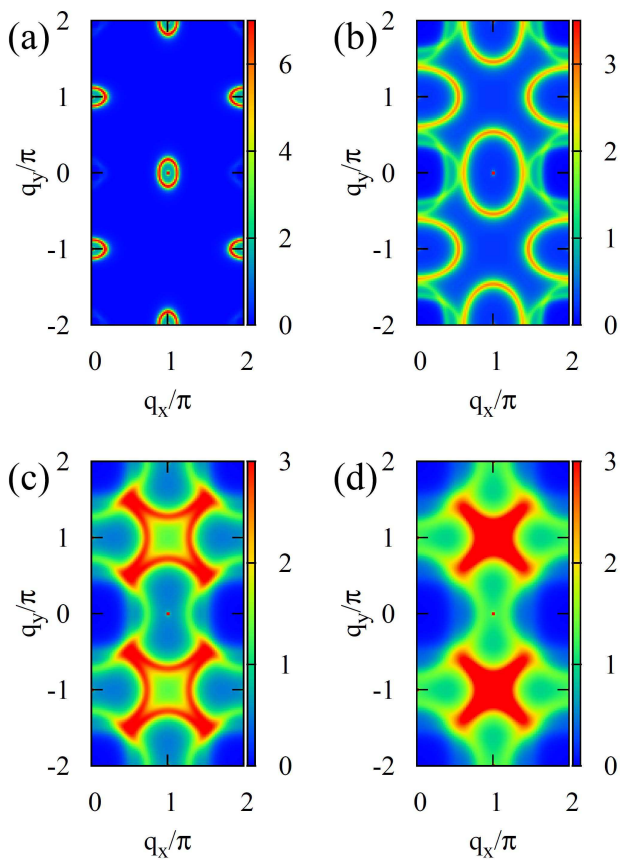}
\caption{(Color online)
Constant energy cuts of the rotational symmetrized spin dynamical structure factor
in the momentum space in the MSW theory for $S=1$, $J_1/J_2=1.0$, $K/J_2=0.8$,
and $T/J_2=1.0$. The corresponding energies are respectively $\omega=4J_2$ in (a), $\omega=10J_2$ in (b),
$\omega=11.5J_2$ in (c), $\omega=12J_2$ in (d). In all the panels, a broadening factor $0.5J_2$ has been used for the convenience of
calculation.
}
\label{fig:Sqw0}
\end{figure}

The dispersion shows a gap at $(\pi,0)$ (and also at $(0,0)$), with the size
\begin{equation}
\Delta_1=\sqrt{-\mu(8J_2g_{x+y}+4\tilde{J}_{1x}g_x-\mu)}. \label{Eq:Gap1}
\end{equation}
At low temperatures the gap is small since $\mu\rightarrow 0$ as $T\ll T_{\sigma0}$.
In this limit, the excitation near $(\pi,0)$ can be approximated by
$\varepsilon_{\mathbf{k}}=\sqrt{v_{1x}^2(\pi-k_x)^2+v_{1y}^2k_y^2+\Delta_1^2}$,
where the velocities are respectively:
\begin{eqnarray}
v_{1x} &=& 4J_2g_{x+y}+2\tilde{J}_{1x}g_x, \label{Eq:v1x}\\
v_{1y} &=& \sqrt{(4J_2g_{x+y}+2\tilde{J}_{1x}g_x) (4J_2g_{x+y}-2\tilde{J}_{1y}f_y)+2\tilde{J}_{1y}f_y\mu}.\nonumber\\
\label{Eq:v1y}
\end{eqnarray}
The excitation develops to the gapless Goldstone mode at $T=0$. At $(\pi,\pi)$ (and also at $(0,\pi)$)
the dispersion has a different gap
\begin{equation}
\Delta_2 = \sqrt{(8J_2g_{x+y}-4\tilde{J}_{1y}f_y-\mu) (4\tilde{J}_{1x}g_x-4\tilde{J}_{1y}f_y-\mu)}. \label{Eq:Gap2}
\end{equation}
The features that $v_{1x}\neq v_{1y}$ and $\Delta_1\neq\Delta_2$ already exists in the isotropic $J_1-J_2$ model.
In the $J_1-J_2-K$ model, $\Delta_1$ at $(\pi,0)$ is only weakly affected by $K$ because it is dominated by $\mu$.
But $\Delta_2$ at $(\pi,\pi)$ is strongly influenced. It increases with $K$. For sufficiently large $K$, approximately
where $\tilde{J}_{1y}$ changes sign to be ferromagnetic (the shaded region in Fig.~\ref{fig:phd2D}), the dispersion
at $(\pi,\pi)$ turns from a local minimum to a maximum, as shown in Fig.~\ref{fig:Dispersions}(a).
Similar behavior in the spin-wave dispersion of the $J_1-J_2-K$ model has also been discussed
in Ref.~\onlinecite{WysockiNP11} in the antiferromagnetically ordered phase, but our results apply to the paramagnetic phase.

\section{Dynamical structure factor}\label{Sec:Sqw}

In order to investigate the magnetic excitations, which are directly accessible by INS measurements,
we have calculated the magnetic structure factor $\mathcal{S}({\mathbf{q},\omega})$. Our main interest is to
understand the experimentally observed anisotropic feature of the magnetic excitations in the paramagnetic phase
above the N\'{e}el temperature. As already discussed in Sec.~\ref{Sec:PhD}, in this temperature regime, the most
relevant factor for the in-plane anisotropy is the Ising order. Therefore, we will concentrate our discussion on the
magnetic structure factor in phase II of the 2D $J_1-J_2-K$ model. In this phase $\mathcal{S}({\mathbf{q},\omega})$
has the same form in both MSW and Schwinger boson theories,
\begin{eqnarray}
\nonumber  \mathcal{S}({\mathbf{q},\omega)} &=& 2\pi\frac{C}{\mathcal{N}} {\sum_{\mathbf{k}}}^\prime \sum_{s,s^\prime=\pm1} \left[ \cosh(2\theta_{\mathbf{k}+\mathbf{q}}-2\theta_{\mathbf{k}}) -ss^\prime\right] \\
& & \times \delta(\omega-s\varepsilon_{\mathbf{k}+\mathbf{q}}-s^\prime\varepsilon_{\mathbf{k}}) n^s_{\mathbf{k}+\mathbf{q}}n^{s^\prime}_{\mathbf{k}},\label{Eq:Sqw}
\end{eqnarray}
where $\sum^\prime$ refers to the summation over the magnetic Brillouin zone corresponding to the $(\pi,0)$ order, which is enclosed by $-\pi/2\leqslant k_x\leqslant \pi/2$, and $-\pi\leqslant k_y\leqslant \pi$.  $n^+_{\mathbf{k}}=n_{\mathbf{k}}+1$ and $n^-_{\mathbf{k}}=n_{\mathbf{k}}$. $C=1$ in the MSW theory, and $C=3/2$ in the Schwinger boson theory with SU($2$) symmetry.\cite{footnote} We see from Eq.~\eqref{Eq:Sqw} that the contribution to $\mathcal{S}({\mathbf{q},\omega})$ comes from two-boson processes. Hence, in general cases the peak of $\mathcal{S}({\mathbf{q},\omega})$ does not follow the boson dispersion. But at low temperatures, the largest contribution to $\mathcal{S}({\mathbf{q},\omega})$ in the summation over $\mathbf{k}$ comes from the term at $\mathbf{k}=(0,0)$ since the small gap $\Delta_1$ at this point results in a large boson number $n_{\mathbf{k}}$. To satisfy the energy conservation in the $\delta$ function, $\mathcal{S}({\mathbf{q},\omega})$ must be peaked at $\omega\approx\varepsilon_{\mathbf{q}}$. Actually this leads to a two-peak structure corresponding to $s^\prime=\pm1$ near
$\omega=\varepsilon_{\mathbf{q}}$ for a fixed $\mathbf{q}$. But the separation of these two peaks is proportional to $\varepsilon_{(0,0)}$ and is very small at low temperatures.
In the numerical calculations performed, the gap between the two peaks
is healed by substituting the delta function by a Lorentzian with a small
broadening width.
As a result of this small broadening, $\mathcal{S}({\mathbf{q},\omega})$ only shows a single peak structure. Therefore, in this limit the peak positions of $\mathcal{S}({\mathbf{q},\omega})$ follow the boson dispersion.

To better discuss the anisotropic distribution of the spectral weight in momentum space, we plot the
constant energy cuts of the calculated $\mathcal{S}(\mathbf{q},\omega)$ at a fixed temperature $T<T_{\sigma0}$ in
Fig.~\ref{fig:Sqw0}. At low energies, the peaks of $\mathcal{S}(\mathbf{q},\omega)$ form a elliptic ring centered
at $(\pi,0)$ (and also its symmetry related point $(0,\pi)$ after rotation symmetrization), as displayed in
Figs.~\ref{fig:Sqw0}(a),(b). The elliptic feature is a consequence of the anisotropic correlation lengths in the
Ising ordered phase, and the ellipticity near $(\pi,0)$ is proportional to $\xi_x/\xi_y=v_{1x}/v_{1y}$,
which is not sensitive to temperature since the mean-field parameters are only weakly temperature dependent
for $T<T_{\sigma0}$ (Fig.~\ref{fig:MFP2D}). The ellipticity
 also only weakly depends on $K$: for $J_1/J_2=1$, we find $\xi_x/\xi_y\simeq 1.7$ at $K=0$, and
 $\xi_x/\xi_y\simeq 1.4$ at $K/J_2=0.8$.

With increasing energy, the ellipse centered around $(\pi,0)$ expands towards the Brillouin zone boundary,
as seen in Figs.~\ref{fig:Sqw0}(a)-(d).
For sufficiently large energy,
the spectral weight reduces greatly along the $q_x$ direction, and the $\mathcal{S}(\mathbf{q},\omega)$
is peaked near $q_y=\pm\pi/2$ along the $q_y$ direction (Fig.~\ref{fig:Sqw0}(c)).
The elliptical peak feature appears to have been split into two parts in the direction of its major axis.
As the energy gets close to $\varepsilon_{(\pi,\pi)}$, the two peaks move towards $(\pi,\pm\pi)$, forming
patterns that are centered around $(\pi,\pm\pi)$; {\it cf.} Figs.~\ref{fig:Sqw0}(c)-(d).
In our theory, there are two factors that contribute to this anisotropic distribution of the spectral weight along the ellipses.
Firstly, for $\omega>\varepsilon_{(\pi/2,0)}$, along the $q_x$ axis the energy conservation
in the $\delta$ function of Eq.~\eqref{Eq:Sqw} can only be satisfied when $\mathbf{k}\neq (0,0)$.
A nonzero $\mathbf{k}$
corresponds to a smaller $n_{\mathbf{k}}$,
which greatly reduces $\mathcal{S}(q_x,\omega)$. Along the $q_y$ axis, however,
$\mathcal{S}(q_y,\omega)$ is not reduced because the
$\mathbf{k}=(0,0)$ mode can still satisfy the energy conservation.
Secondly, for a given $\kk$, the coherence factor $\cosh(2\theta_{\mathbf{k}}-2\theta_{\mathbf{k}+\mathbf{q}})$
along the ellipse is also anisotropic.
To see this, recall that the largest contribution to $\mathcal{S}(\mathbf{q},\omega)$ is from the $\kk=(0,0)$ term
in Eq.~\eqref{Eq:Sqw}. For simplicity,
we take a single mode approximation, namely, $\mathcal{S}(\mathbf{q},\omega)$ can be approximated by
this $\kk=(0,0)$ term.
Then the ellipse showing spectral weight peaks is determined by $\varepsilon_\mathbf{q}=\omega$, and the coherence factor
$\cosh(2\theta_{\mathbf{k}}-2\theta_{\mathbf{k}+\mathbf{q}})\propto (B_\mathbf{q}-A_\mathbf{q})/\varepsilon_\mathbf{q}$.
For $\sqrt{q_x^2\xi_x^2+q_y^2\xi_y^2} \ll T/\Delta_1$,
$B_\mathbf{q}-A_\mathbf{q}\approx \Delta-\varepsilon_\mathbf{q}/\Delta-2\tilde{J}_{1y}f_yq_y^2$, where
$\Delta=8J_2g_{x+y}+4\tilde{J}_{1x}g_x$. Since $\tilde{J}_{1y}<0$ for the choice of model parameters, it is easy to see
that along the ellipse $\varepsilon_\mathbf{q}=\omega$, the maximum of the coherence factor is located along the $q_y$ axis
but not the $q_x$ axis. Since within the single mode approximation, $\mathcal{S}(\mathbf{q},\omega)$ is proportional
to the coherence factor, $\mathcal{S}(\mathbf{q},\omega)$ is also anisotropic along the ellipse. Note that at low energies
($\sqrt{q_x^2\xi_x^2+q_y^2\xi_y^2} \ll T/\Delta_1$), $\tilde{J}_{1y}f_yq_y^2\ll \Delta$, so the anisotropy is very small.
This coherence-factor-induced anisotropy becomes sizable when the ellipse is large
(for $\sqrt{q_x^2\xi_x^2+q_y^2\xi_y^2} \gtrsim T/\Delta_1$).

\section{Discussions}\label{Sec:Disc}

\subsection{The effects of spin size}
\begin{figure} [h]
\centering\includegraphics[scale=0.3]{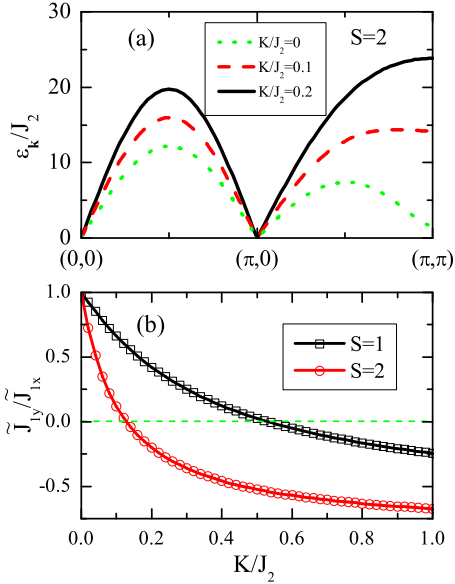}
\caption{(Color online)
(a): MSW dispersion along high symmetry directions in the
paramagnetic Brillouin zone for the 2D $J_1-J_2-K$ model at $S=2$,
$J_1/J_2=1.0$, $T/J_2=1.0$, and various $K$ values.
(b): The ratio of $\tilde{J}_{1y}/\tilde{J}_{1x}$, showing the anisotropy in effective exchange couplings, see Eq.~(\ref{Eq:Jxy}), as a function of the biquadratic coupling for $S=1$ and $S=2$.
}
\label{fig:S1S2comp}
\end{figure}
Besides the $S=1$ results shown in Sec.~\ref{Sec:PhD} and Sec.~\ref{Sec:Sqw}, we have also studied the $J_1-J_2-K$ model with larger spin sizes, and found the mean-field phase diagram is similar to the one
in Fig.~\ref{fig:phd2D}. Approximately, $T_{\sigma0}$ and $T_0$ are increased by a factor of $S(S+1)/2$. The boson dispersion shown in Fig.~\ref{fig:S1S2comp}(a) also exhibits the similar features as in the $S=1$ case. In Fig.~\ref{fig:S1S2comp}(b) we compare the ratio of the effective nearest neighbor couplings $\tilde{J}_{1y}/\tilde{J}_{1x}$, defined in Eq.~(\ref{Eq:Jxy}), for $S=1$ and $S=2$ (at zero temperature). We find that with increasing $S$, the minimal $K/J_2$ value where $\tilde{J}_{1y}$ becomes ferromagnetic is dropped from $K/J_2\approx0.53$  to $K/J_2\approx0.13$. Hence, we conclude that the anisotropy of the effective exchange couplings induced by non-Heisenberg coupling $K$ is more significant for larger spin size $S$.

We can further compare our MSW result at $T=0$ with the one in a recent MSW study,
which used a mean-field treatment that is different from ours.\cite{Stanek11} The
 two theories yield exactly the same results when the spin size $S\to\infty$.
For finite spin sizes, by comparing the behavior of $\tilde{J}_{1y}/\tilde{J}_{1x}$ ratio in Fig.~\ref{fig:S1S2comp}(b) and the corresponding
results in Ref.~\onlinecite{Stanek11}, we observe that the two theories give qualitatively similar results
for the anisotropy in the exchange couplings:
The biquadratic coupling $K$ reduces the ratio of the effective ratio $\tilde{J}_{1y}/\tilde{J}_{1x}$.
Quantitatively, there are some differences between the two approaches. In particular, while for $S >1$, the ratio $\tilde{J}_{1y}/\tilde{J}_{1x}$
changes sign at a finite $K$ value in both theories, this sign change does not appear for $S=1$ in Ref.~\onlinecite{Stanek11}.

\subsection{Generalizations of the $J_1-J_2-K$ model}
Several remarks on the $J_1-J_2-K$ model studied in this paper.
From the incipient Mott picture, when the system is in the vicinity of $U_c/t$, the spin Hamiltonian
contains interactions involving more than just two spins.
To see this, we start from a multi-orbital Hubbard model on a square
lattice, and assume that Hund's rule coupling locks the spins in different
orbitals to a high spin state. Then we may obtain a spin-only
Hamiltonian by integrating out the fermion degrees of freedom based
on perturbation  in $t/U$. To the $t^2/U$ order,
we obtain the usual $J_1-J_2$ Heisenberg interaction between
nearest- and next-nearest- neighbor spins.
The next-order terms appear in the order $t^4/U^3$, and include the biquadratic $K$ term as well as the ring exchange interactions. Here, we have focused on the effects of the biquadratic interaction. The influence of the ring exchange interactions in the regime we are considering is briefly discussed in Appendix ~\ref{ringexchange}.

To fully understand the antiferromagnetic $(\pi,0,\pi)$ order revealed in the experiments, the 2D $J_1-J_2-K$ model
needs to be extended to the 3D case by including an interlayer coupling $J_z$.
A nonzero $J_z$ will support the antiferromagnetic order up to the N\'{e}el temperature $T_N$. In the mean-field treatment,
the antiferromagnetic order emerges at a mean-field N\'{e}el temperature $T_{N0}$. The details of the effects
of the interlayer coupling $J_z$ to the magnetic phase diagram of the $J_1-J_2-K$ model and
the magnetic excitation spectrum is further discussed in Appendix ~\ref{Sec:PhD3D}.

When fluctuations beyond the mean-field level are taken into account, the actual N\'{e}el and Ising transition
temperatures, $T_N$ and $T_{\sigma}$ can be well below their mean-field values. The mean-field temperatures
$T_{N0}$ and $T_{\sigma0}$ then correspond to some crossover temperature scales, below which the fluctuating order
have significant effects.
The fluctuating anisotropic
effects we have presented will be dominant
in the temperature regime $T_{N}<T<T_{\sigma0}$.\cite{Goswami11}

\subsection{Effect of itinerant electrons}\label{Sec:damping-fermion}

\begin{figure} [h]
\centering\includegraphics[scale=0.3]{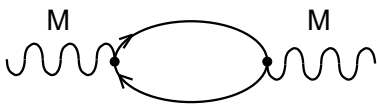}
\caption{(Color online)
Diagramm of the second-order contribution to the effective action in Eq.~\eqref{Eq:action} due to coupling to fermions.
}
\label{fig:bubble}
\end{figure}

Within the bad-metal description of the iron pnictides, the quasi-localized moments are coupled to itinerant electrons with a
spectral weight that depends on the proximity to the Mott transition. A convenient way to describe the effect of
itinerant electrons on the spin dynamics is to reformulate the results of the local-moment-based calculations in terms
of a non-linear sigma model, and introduce into the latter a damping caused by the itinerant electrons; for details,
we refer to Ref.~\onlinecite{Goswami11}. Well below $T_{\sigma0}$ and in the vicinity of $(\pi,0)$, the effects of the itinerant electrons are described in terms of the effective action for the staggered magnetization ${\bf M}$:
\begin{eqnarray} \label{Eq:action}
S[M] &=& T\int d {\bf q} \sum_{l}\left[r + \Delta r + v_{1x}^2 q_x^2 + v_{1y}^2 q_y^2 + \omega_l^2\right.\nonumber\\
 && \left. + {\gamma}|\omega_l|\right] {\bf M}^2
 + u {\bf M}^4 + \mathcal{O}({\bf M}^6).
\end{eqnarray}
Here, ${\bf M} =\mathbf{m}+\mathbf{m}^{'}$ is the sum of ${\bf m}$ and ${\bf m '}$, the O(3) vectors respectively for the magnetizations of the two decoupled sublattices on the square lattice, and $\omega_l$ the Matsubara frequency.
This action arises in a ``$w$-expansion", which is based on a proximity to the Mott transition and is described in Refs.~\onlinecite{Si_NJP,Dai}; it
has the form of the usual $\sigma$-model~\cite{nagaosa-book}.
 In the first term,
$\Delta r>0$ is a mass shift and $\gamma$ describes the strength of spin damping from coupling to fermions. (See Fig.~\ref{fig:bubble})
 At relatively low energies, this introduces a procedure that can be used
to describe the broadening of the spin spectral peaks in momentum space due to coupling to itinerant electrons.\cite{Goswami11}

We should emphasize that this procedure is a qualitative treatment of the spin damping. Incorporating the
full details of the electronic bandstructure will introduce momentum-dependence of the damping rate, making it
possible to generate the type of anisotropic damping that
was proposed phenomenologically by Harriger {\it et al.} \cite{Harriger}.

Comparing our results for the $J_1-J_2-K$ model in Fig.~\ref{fig:Sqw0}(a)-(d) with those of the $J_1-J_2$ model (Fig.~4 of
Ref.~\onlinecite{Goswami11}) shows that, the biquadratic term itself brings out an anisotropy in the spectral weight of the
elliptic peaks. The spectral weight is larger along the major axis of the ellipse than along its minor axis.
This anisotropy goes in the same direction as that of the experimental data on BaFe$_2$As$_2$,
illustrated in Fig.~\ref{fig:Sqw0}(f).
We therefore conclude that both the ellipticity and intensity anisotropy of the spectral peaks in momentum space
are controlled by the exchange interactions.

We note that the Ising order parameter is also coupled to the itinerant electrons. Since the Ising order parameter
breaks the $C_{4v}$ symmetry, it couples to  those spin singlet fermion bilinears that
 correspond to the $B_{1g}$ representation. Consequently a nonzero Ising order parameter will induce a nonzero
 $d_{x^2-y^2}$ nematic charge density $(\cos k_x-\cos k_y)c_{\mathbf{k},\alpha}^{\dagger}c_{\mathbf{k},\alpha}$
 for all the d-orbital electrons, where $\alpha$ is the orbital index. In addition the Ising order parameter will induce
 a nonzero charge density imbalance
 $c_{\mathbf{k}xz}^{\dagger}c_{\mathbf{k}xz}-c_{\mathbf{k}yz}^{\dagger}c_{\mathbf{k}yz}$ between
 the $d_{xz}$ and the $d_{yz}$ orbitals, which is also referred as the ferro-orbital order. As a result the
 spin fluctuations from the incoherent degrees of freedom can give rise to an orbitally ordered, charge nematic metal,
 with anisotropic transport properties. In a model with 3D
coupling (see Appendix~\ref{Sec:PhD3D}), the coupling to the itinerant electrons will reduce the N\'{e}el
transition temperature from its
mean-field value $T_{N0}$ to $T_N < T_{N0}$, through the positive $\Delta r $ noted above.
It will likewise decrease the
 Ising transition temperature from its mean-field value $T_{\sigma0}$ to $T_{\sigma} < T_{\sigma0}$.
 However, the correlation lengths are still sizeable
 and should be anisotropic up to $T_{\sigma0}$.\cite{Goswami11} This implies that, in the 3D model with
 three-dimensional coupling, we expect anisotropic magnetic excitations to exist from $T_N$ all the way up to the crossover
 temperature scale $T_{\sigma0}$, in the absence of a static Ising order.

\subsection{Comparison with other approaches} \label{Sec:other}

Our studies in the $J_1-J_2-K$ model, with or without the coupling to the itinerant electrons, are very different from purely itinerant
studies with $U/t$ much smaller than $U_c/t$. Because the Fermi surface comprises small electron and hole pockets, such calculations are expected to yield very small spin spectral weight. Experimentally, the total spectral weight is known
to be large, with an effective moment that is larger than $1$ $\mu_B$/Fe in CaFe$_2$As$_2$
(Ref.~\onlinecite{Zhao}). Such a large spectral weight arises
naturally in our approach using as the starting point the $J_1-J_2$ model (with or without  the $K$ term).

Our approach can be compared more closely with that of the DMFT studies
of Ref.~\onlinecite{ParkHauleKotliar11}, in which the ratio of the
 effective interaction (combined Coulomb and Hund's interactions)
to the characteristic bandwidth is close to the Mott-transition value,
 $U/t \lessapprox U_c/t$.
 The proximity to the Mott transition
 ensures
 that a large part of the electronic spectral weight lies in the incoherent regime,
 which will naturally give rise to a large spin
spectral weight.
 The consistency of the momentum-dependence determined by the DMFT calculations and
 that of our $J_1-J_2-K$ calculations further suggests the compatibility of the two approaches.
 There are some important
 differences, however. In the DMFT calculation, the anisotropy of the structure factor has been attributed
 to the geometry of the Fermi surface(s). The $J_1-J_2-K$ results however tie the anisotropy of the spin spectral weight
in  momentum space with
 the Ising correlations.

Experimentally, the Ising correlations can be very naturally connected with the $x-y$ anisotropy observed
in ARPES\cite{Yi} and transport\cite{Fisher} measurements in the detwinned
122 iron pnictides at temperatures above $T_N$. A recent theoretical calculation\cite{Fernandes} shows
how resistivity anisotropy in the tetragonal phase above $T_N$ follows
from the existence of the Ising correlations discussed here.

\begin{figure*} [t!]
\centering
\includegraphics[
width=140mm
]{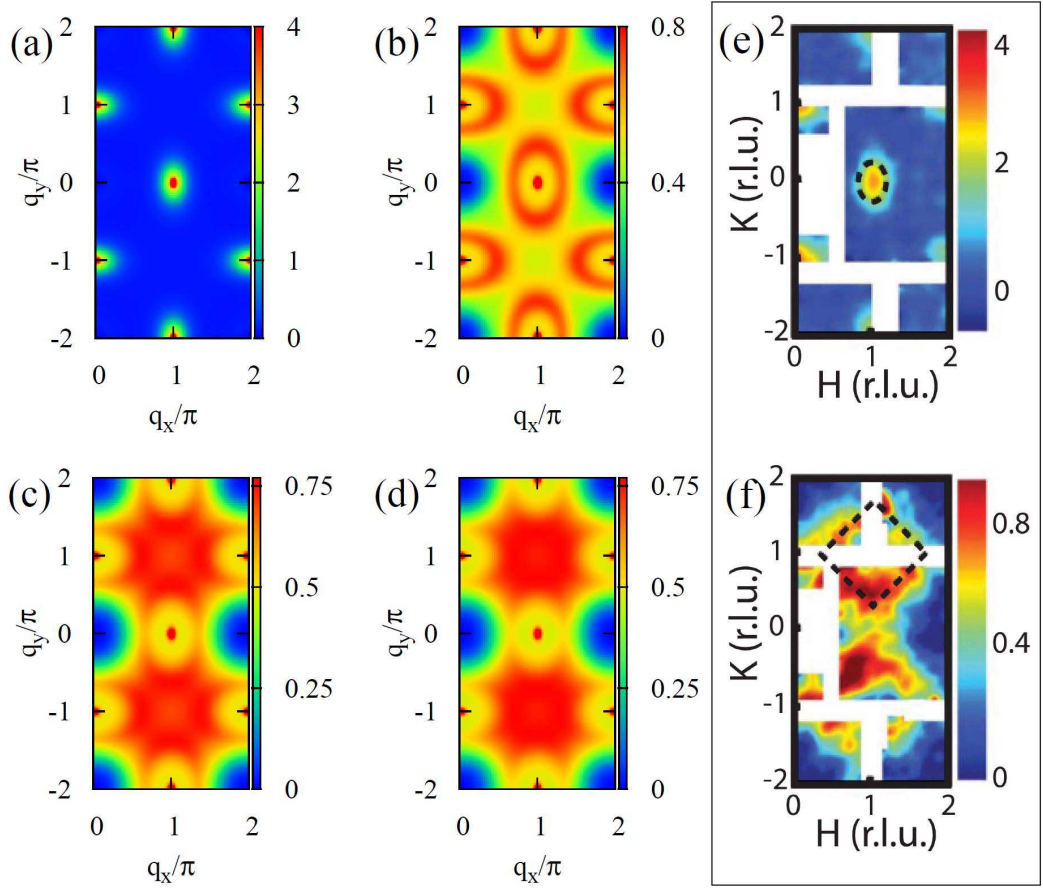}
\caption{(Color online)
Evolution of $\mathcal{S}(\mathbf{q},\omega)$ in the paramagnetic phase of the $J_1-J_2-K$ model,
showing that the elliptical features near $(\pi,0)$
at low energies (top panels) are split into features that are centered around $(\pi,\pi)$, as the energy is increased towards
the zone-boundary spin-excitation energy (bottom panels).
This trend is consistent with the inelastic neutron scattering experiments,
shown in the box on the right for two energies measured in the paramagnetic phase of BaFe$_2$As$_2$ (data taken from Ref.~\onlinecite{Harriger}).
(a)-(d): Same as Fig.~\ref{fig:Sqw0}(a)-(d), but with damping $\gamma=3J_2$. (e)-(f): The INS data at $T=150$ K in BaFe$_2$As$_2$, taken from Ref.~\onlinecite{Harriger}. The energy transfer is
$\omega=50\pm10$ meV in (e), and $\omega=150\pm10$ meV in (f). Here we find that the best agreement between theory and experimental data achieves when taking $J_2\approx13$ meV in the model.
}
\label{fig:Sqwcomp}
\end{figure*}
\section{Comparison with experiments on the paramagnetic phases of parent 122 iron pnictides}\label{Sec:expt}

Spin dynamics in the paramagnetic phase of the parent 122 iron pnictides has been recently studied
via INS measurements.\cite{Diallo,Harriger,Ewings} For CaFe$_2$As$_2$, spin dynamics
at low energies (below $70$ meV) has been studied by Diallo {\it et al}.\cite{Diallo} It is found that
the peaks of the dynamical structure factor form anisotropic elliptic features at low energies,
similar to the results in the antiferromagnetic phase.\cite{Zhao} More recently,
Harriger {\it et al.}\cite{Harriger} measured the spin dynamics
of BaFe$_2$As$_2$ up to $200$ meV.  At low energies, they found the distribution of spectral weights
in the momentum space forms similar elliptic feature as in the CaFe$_2$As$_2$ case. With increasing energy,
the elliptic feature expands towards the Brillouin zone boundary.
Moreover, they determined the magnetic dispersion to be peaked (or flat-topped) near
$(\pi,\pi)$.
Similar results have also been reported for SrFe$_2$As$_2$.\cite{Ewings}

Our study on the $J_1-J_2-K$ model have already provided valuable information for understanding these
experimental observations. In real materials, the various fluctuation mechanisms and the coupling to fermions/phonons
will reduce the N\'{e}el and Ising transition temperatures. However, below the mean-field Ising temperature $T_{\sigma0}$, the effective couplings between
the nearest neighbors are always anisotropic. Hence we expect the magnetic fluctuations
to be anisotropic for $T_N (\le T_{\sigma})<T<T_{\sigma0}$,
which corresponds to the upper portion of region II in Fig.~\ref{fig:phd2D}.
This anisotropy is reflected in the spin dynamics in the paramagnetic phase.

To be specific, the anisotropic elliptic feature at low energies observed in CaFe$_2$As$_2$ and
other parent 122 compounds can already been understood within the $J_1-J_2$ model.\cite{Goswami11}
We have shown in Fig.~\ref{fig:Sqw0} that the $J_1-J_2-K$ model gives the similar low-energy elliptic feature.
It will be important to measure the spin dynamics at high energies in this material.

Our calculated evolution of this elliptic feature as the energy is raised in the $J_1-J_2-K$ model
can be systematically compared
with the experimental  observations in BaFe$_2$As$_2$
and SrFe$_2$As$_2$. To see this, we fit the peak positions of calculated
$\mathcal{S}({\mathbf{q},\omega})$ to the experimental magnetic excitation
dispersion data in BaFe$_2$As$_2$, from which we can extract the best fitted
values of the exchange couplings. Assuming $S=1$, we find the fitted exchange
couplings are $J_2=17\pm4$ meV, $J_1/J_2=1.0\pm0.5$, and $K/J_2=0.8\pm0.3$.
We find that a very broad range of the $J_1/J_2$ ratio can all fit the experimental data quite well.
As illustrated in Fig.~\ref{fig:Dispersions}(b), any dispersion curve within the shaded region
fits the experimental data within error bars. But to fit the dispersion data near the local maximum at $(\pi,\pi)$,
a moderate $K/J_2$ ratio is necessary. For $S=1$, we find $K/J_2\approx0.8$ fits the data the best.
On the other hand, for $S=2$, the best fitted $K/J_2$ ratio is substantially reduced to about $0.2$.

For BaFe$_2$As$_2$, detailed measurements in the momentum space have been reported
by Harriger {\it et al.} \cite{Harriger}. This allows us to see that
the agreement between our theory and the experiment is not only for the dispersion,
but also for the anisotropic distribution of the spectral weight of $\mathcal{S}(\mathbf{q},\omega)$ in momentum space.

In order to make a comparison with experimental data, we use Eq.~(\ref{Eq:Sqw}) in the calculation of $\mathcal{S}(\mathbf{q},\omega)$ and approximate the delta function by the following Lorentzian broadening
\begin{equation} \label{Eq:lorentzian}
\delta(\omega-\Delta \varepsilon) \longrightarrow \frac{1}{\pi}\frac{\gamma}{(\omega - \Delta \varepsilon)^2 + \gamma^2}.
\end{equation}
Here we have assumed that the broadening mainly comes from the damping effect due to coupling to itinerant electrons. It is then reasonable to take the phenomenological broadening factor to be the damping $\gamma$ introduced in Eq.~\eqref{Eq:action} since in either the MSW or Schwinger boson theory, the damping is still due to the same bubble in Fig.~\ref{fig:bubble}. Calculating the magnitude of $\gamma$ requires a detailed microscopic theory and is beyond the scope of this article, however we can use Ref.~\onlinecite{Goswami11} for reference, where it has been determined that $\gamma/J_2\approx3$ for CaFe$_2$As$_2$. Here we assume that this ratio still holds for BaFe$_2$As$_2$ and the damping is isotropic. In Figs.~\ref{fig:Sqwcomp}(a)-(d) we replot the theoretical dynamical spin structure factor in Fig.~\ref{fig:Sqw0} with this damping factor, and compare them with the experimental data in Ref.~\onlinecite{Harriger}.
At low energies, our theory correctly captures the elliptical feature centered at $(\pi,0)$
as displayed in Figs.~\ref{fig:Sqwcomp}(a),(b). Experimentally, this is seen as a filled elliptical spot due to damping effect, which is also shown in our theoretical plot in Fig.~\ref{fig:Sqwcomp}(a).
The evolution of the elliptical feature with increasing energy is also consistent with the experimental observation:
as the ellipse expands towards zone boundary, it gradually splits into two parts, and forms a pattern around $(\pi,\pi)$
(see Figs.~\ref{fig:Sqwcomp}(c), (d), and (f)). We reiterate that such anisotropic features are the properties of our $J_1-J_2-K$ model
either with an isotropic or without additional damping due to itinerant electrons. While anisotropic damping proposed in
Ref.~\onlinecite{Harriger} could reinforce the effect, it is not necessary to understand the INS experiments. In CaFe$_2$As$_2$ the elliptical feature around $(\pi,0)$ persists up to high energies, while in BaFe$_2$As$_2$, this elliptical feature splits into two parts at intermediate energy.~\cite{Harriger} These two different behaviors can both be understood within our $J_1-J_2-K$ model with similar, nearly isotropic damping but different $K$ values.

\section{Conclusions}\label{Sec:Conclusion}
In this paper we have investigated the finite
temperature spin dynamics of a $J_1-J_2-K$ antiferromagnetic Heisenberg model using both MSW and SBMF theories. The spin dynamics obtained from these two methods are similar to each other.

We have found that by including a moderate biquadratic coupling $K$, the magnetic excitation spectrum of the $J_1-J_2-K$ model is anisotropic below a mean-field Ising transition temperature $T_{\sigma0}$.
As in the case of the  $J_1-J_2$ model \cite{Goswami11},
the peak of the low-temperature dynamical
structure factor $\mathcal{S}(\mathbf{q},\omega)$
contains elliptical features near $(\pi,0)$ in the paramagnetic Brillouin zone at low excitation energies.
However, unlike the pure  $J_1-J_2$ model, the spectral intensity also displays anisotropy along the ellipse,
with the intensity being higher along the major axis than that along the minor axis.
This spectral anisotropy accounts for the observed particular way in which the low-energy elliptical features, centered around
$(\pi,0)$, expand towards the zone boundary as the energy is increased towards the zone-boundary
spin-excitation energy. It also gives rise to a particular form of
 high-energy spectral features that are centered around $(\pi,\pi)$.

We have also compared our calculated dynamical spin structure factor
of the $J_1-J_2-K$ model with the recent inelastic neutron-scattering measurements
in the paramagnetic phases of the 122 iron pnictides \cite{Harriger,Diallo,Ewings}.
The theoretical results
provide a very natural understanding of the salient features of the experiments.

\acknowledgements

We thank P. Dai, R. A. Ewings and R. Fernandes for
useful discussions, and
NSF Grant No. DMR-1006985 and the Robert A. Welch Foundation
Grant No. C-1411 for partial support. P. G. was supported at the National High Magnetic Field Laboratory by NSF Cooperative Agreement No. DMR-0654118, the State of Florida, and the U. S. Department of Energy. Part of this work was carried out at the Aspen Center for Physics (NSF grant 1066293).

\appendix

\section{Ising transition at small $J_1/J_2$ ratios}\label{Sec:IsingTransition}

\begin{figure} [h!]
\centering\includegraphics[scale=0.3]{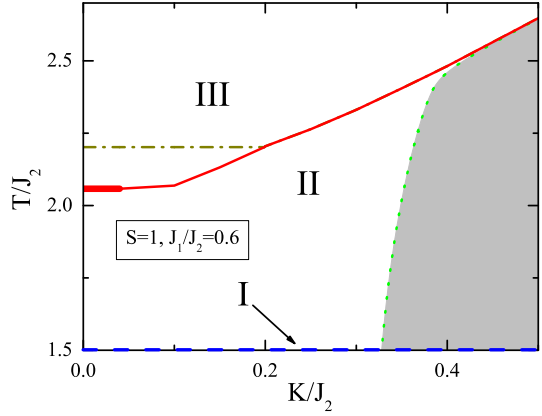}
\caption{(Color online)
Mean-field magnetic phase diagram in the MSW theory for $S=1$, $J_1/J_2=0.6$.
The dashed blue and dashed dotted brown curves refer to the mean-field temperature scales $T_{N0}$ and $T_0$,
respectively. The thicker solid red curve refers to a second-order Ising transition at $T_{\sigma0}$, while the thinner
solid red curve refers to a first-order transition. In the shaded region, the effective exchange coupling $\tilde{J}_{1y}<0$.
}
\label{fig:phd2DsJ1}
\end{figure}

We find that the nature of the mean-field Ising transition at $T_{\sigma0}$ depends on both
$J_1/J_2$ and $K/J_2$ ratios. At $K=0$ and $J_1/J_2\lesssim0.9$, we find $T_{\sigma0}<T_0$,
and the Ising transition at $T_{\sigma0}$ is always second-order (Fig.~\ref{fig:phd2DsJ1}).
When $J_1/J_2\gtrsim0.9$, $T_{\sigma0}$ meets $T_0$ and the Ising transition becomes first-order.
This is an artifact of the mean-field approximation since the transition at $T_0$ is always first-order.\cite{Takahashi1,Takahashi2}
Still for $J_1/J_2\lesssim0.9$, increasing $K$ from zero,  the transition at $T_{\sigma0}$ changes from second-order to first-order
when $K$ is bigger than a bicritical point $K_c$. As shown in Fig.~\ref{fig:phd2DsJ1} for $J_1/J_2=0.6$,
$K_c/J_2\approx0.04$. At $K\gtrsim K_c$, $T_{\sigma0}<T_0$. This suggests that the Ising transition near $K_c$
is not influenced by $T_0$ , but the order of this transition is tuned by $K$. Hence the first-order transition
at $T_{\sigma0}$ is \emph{not} an artifact of the mean-field treatment.

\section{Effects of ring exchange couplings}\label{ringexchange}
Besides the quadratic and biquadratic interactions, other interactions involving
more than two spins can also appear in the spin Hamiltonian in the vicinity of Mott transition.
For instance, the four-spin ring exchange interaction can appear as a consequence of the fourth-order
perturbation associated with the electron hopping process.
We can consider the effects of a four-spin ring exchange process on the spin dynamics by adding
a term
$K_{\square}\sum_{i,j,k,l}[(\mathbf{S}_i\cdot\mathbf{S}_j)
(\mathbf{S}_k\cdot\mathbf{S}_l) -(\mathbf{S}_i\cdot\mathbf{S}_k)(\mathbf{S}_j\cdot\mathbf{S}_l)
+(\mathbf{S}_i\cdot\mathbf{S}_l)(\mathbf{S}_j\cdot\mathbf{S}_k)]$ to the Hamiltonian,
where $K_{\square}>0$, and the sites $(i,j,k,l)$ are the vertices of a square plaquette, labeled clockwise.
The four spin ring exchange competes against $J_{1}$ and $J_2$ and tends to weaken the antiferromagnetic order
coming from $J_1$ or $J_2$. In the linear spin wave description of the $(\pi,0)$ ordered state, we obtain
the effective exchange constants $\tilde{J}_{1x}=J_1+2(K-K_{\square})S^2$, $\tilde{J}_{1y}=J_1-2(K-K_{\square})S^2$,
and $\tilde{J}_{2}=J_2+K_{\square}S^2$, and a reduced spin gap at $(\pi, \pi)$. This trend also persists
in the paramagnetic state, and reduces the size of the Ising order parameter. For consistency with
the experimental results we require  $K>K_{\square}$.

\section{Effects of interlayer exchange coupling}\label{Sec:PhD3D}

\begin{figure} [h]
\centering\includegraphics[scale=0.3]{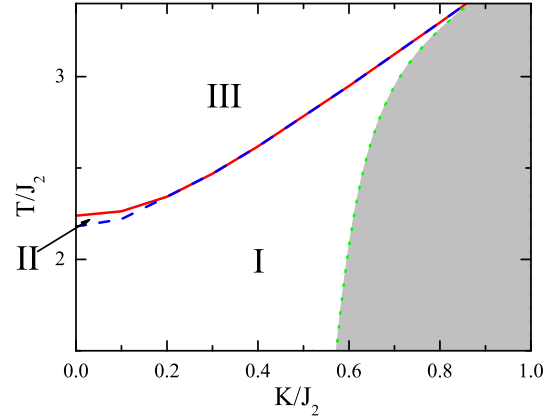}
\caption{(Color online)
Mean-field magnetic phase diagram in the MSW theory for $S=1$, $J_1/J_2=1$, and
an interlayer exchange coupling $J_z/J_2=0.1$. The dashed blue and solid red curves refer to the mean-field
temperature scales $T_{N0}$ and $T_{\sigma0}$, respectively. In the shaded region, the effective exchange
coupling $\tilde{J}_{1y}<0$.
}
\label{fig:phd3D}
\end{figure}

\begin{figure} [h]
\centering\includegraphics[scale=0.325]{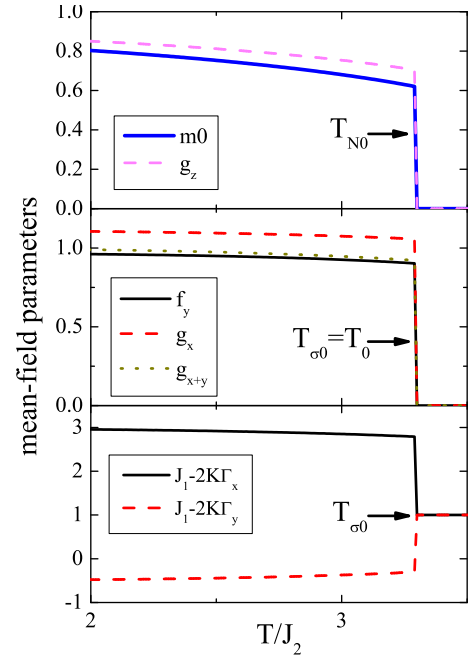}
\caption{(Color online)
The temperature evolution of the mean-field parameters in the MSW theory for $S=1$,
$J_1/J_2=1$, $K/J_2=0.8$, and with an interlayer exchange coupling $J_z/J_2=0.1$.
 }
\label{fig:MFP3D}
\end{figure}

\begin{figure} 
[h]
\centering\includegraphics[scale=0.3]{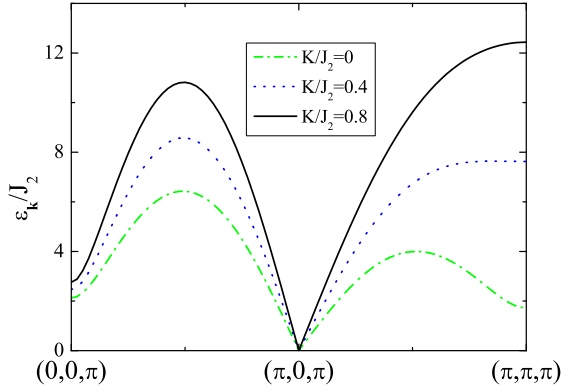}
\caption{(Color online)
Dispersion of the $J_1-J_2-K$ model in the MSW theory for various $K$ values at $S=1$, $J_1/J_2=1.0$,
$T/J_2=0.1$, and with an interlayer exchange coupling $J_z/J_2=0.1$.
}
\label{fig:Dispersions3D}
\end{figure}

The real materials have a 3D tetragonal structure. In the $J_1-J_2-K$ model, the 3D effects can be studied
by extending the model to include a finite interlayer exchange interaction
$J_z\sum_{i}\mathbf{S}_{i}\cdot \mathbf{S}_{i+\hat{z}}$. In 3D the long-range antiferromagnetic phase survives
at finite temperature up to the N\'{e}el temperature $T_{N}$. In MSW and
SBMF theories, the mean-field N\'{e}el temperature $T_{N0}$ is determined by the onset
of spontaneous sublattice magnetization $m_0$. In general, $T_{N0}\leqslant T_{\sigma} \leqslant T_0$. The modification
to our discussion in Sec.~\ref{Sec:Model} comes through an additional interlayer antiferromanetic bond
correlation parameter $g_z$. In the presence of $J_z$, the self-consistent equations of
Eqs.~\eqref{Eq:MSWSelf-Consist1}-\eqref{Eq:MSWSelf-Consist3} and
Eqs.~\eqref{Eq:SBSelfConsistEq1}-\eqref{Eq:SBSelfConsistEq3} are unchanged, but the expressions
for $A_{\mathbf{k}}$ and $B_{\mathbf{k}}$ are modified according to
\begin{eqnarray}
A_{\mathbf{k}}^{\mathrm{3D}} &=& A_{\mathbf{k}}+2J_z g_z \cos k_z\\
B_{\mathbf{k}}^{\mathrm{3D}} &=& B_{\mathbf{k}}+2J_zg_z,
\end{eqnarray}
in the MSW mean-field theory, and
\begin{eqnarray}
A_{\mathbf{k}}^{\mathrm{3D}} &=& A_{\mathbf{k}}+2J_z g_z \sin k_z\\
B_{\mathbf{k}}^{\mathrm{3D}} &=& B_{\mathbf{k}}.
\end{eqnarray}
in the SBMF theory.

In Fig.~\ref{fig:phd3D} we show the phase diagram at the experimentally suggested ratio $J_z/J_2=0.1$. Similar to the
2D case, the mean-field phase diagram consists of an Ising and N\'{e}el ordered antiferromagnetic phase (I),
an Ising ordered but N\'{e}el disordered paramagnetic phase (II), and an Ising and N\'{e}el disordered paramagnetic
phase (III), separated by mean-field temperatures $T_{N0}$ and $T_{\sigma0}$ (see also Fig.~\ref{fig:MFP3D}).
For the parameters in Fig.~\ref{fig:phd3D}, the transitions are both first-order, and both $T_{N0}$ and $T_{\sigma0}$ increase with $K$. For $K/J_2\gtrsim 0.2$,
$T_{N0}$ meets $T_{\sigma0}$, and there is only a single transition between phases I and III.
The absence of phase II in this regime is an artifact of the mean-field theory, since $T_{\sigma0}$ is always
bounded above by the mean-field scale $T_0$.

In connection to the real materials, we note that the structural and magnetic transitions in the 1111 pnictides are well separated. But in 122 compounds, they are either very close to each other, or become a single first-order transition. This can be understood in terms of the present theory, provided $J_z$ is stronger in the 122 materials. By comparing Fig.~\ref{fig:phd3D} and Fig.~\ref{fig:MFP2D} we see that the magnetic transition is closer to the Ising transition for a larger $J_z$. Recent experiments also show that the electron doping may cause the separation of the structural and magnetic transition temperatures in Ba(Fe,Co)$_2$As$_2$ system.\cite{Rotundu11} The similarity between this behavior and the $K$ dependence of $T_{\sigma0}$ and $T_{N0}$ in the phase diagram of Fig.~\ref{fig:phd3D} suggests the possibility that electron doping is
positively correlated with a reduction of the biquadratic
interaction. It would then be interesting to reveal the link between them in future experimental and theoretical studies.

In Fig.~\ref{fig:Dispersions3D} we show the low-temperature boson dispersions of the 3D model for various $K$
values along two high-symmetry directions in the $k_z=\pi$ plane. Aside from a larger gap at $(0,0,\pi)$,
the dispersion is very similar to the one in 2D: the dispersion is highly anisotropic, and with increasing $K$,
the local minimum at $(\pi,\pi,\pi)$ turns to a maximum. This is not too surprising because the in-plane anisotropy
is a consequence of the 2D Ising-type fluctuations, and is not sensitive to the interlayer exchange coupling.


\begin{thebibliography}{99}
\bibitem{footnote} The factor $C=(N+1)/N$ in a $\rm{SU}(N)$ or $\rm{Sp}(N)$ SBMF theory. Hence $\mathcal{S}({\mathbf{q},\omega})$ satisfies the sum rule only when
$N\rightarrow\infty$. In the $\rm{SU(2)}$ SBMF theory, $C=3/2$ violating the sum rule
is a known issue. See Ref.~\onlinecite{Arovas} for more detail.
\end{thebibliography}
\end{document}